\documentclass[paper,twocolumn,showpacs,superscriptaddress,prl]{revtex4}%
\usepackage{amsfonts}
\usepackage{amsmath}
\usepackage{amssymb}
\usepackage{graphicx}%
\begin{document}
\title{Effects of optical fields on the tunneling time of chiral electrons in graphene}
\author{Jiang-Tao Liu}
\email[Electronic address:]{jtliu@semi.ac.cn}
\affiliation{Department of Physics, Nanchang University, Nanchang
330031, China}%
\author{Fu-Hai Su}
\affiliation{Key Laboratory of Materials Physics, Institute of Solid
State Physics, Chinese Academy of Sciences, Hefei 230031, China}
\author{Hai Wang}
\affiliation{The Beijing Key Laboratory for Nano-Photonics and Nano-Structure, Department of Physics, Capital Normal University,
Beijing 100037, China}%
\author{Xin-Hua Deng}
\affiliation{Department of Physics, Nanchang University, Nanchang 330031, China}%

\date{\today}

\begin{abstract}
The influences of optical fields on the tunneling time in graphene are investigated in real time using the finite-difference time-domain method. The tunneling time of electrons irradiated by an optical field is significantly different from that observed in traditional quantum tunneling. We found that when the barrier width increases, the group delay becomes constant for the reflected wave packet, but increases linearly for the transmitted wave packet. This peculiar
tunneling effect can be attributed to current leakage in a
time-dependent barrier generated via the optical Stark effect.

\end{abstract}
\maketitle


Quantum tunneling time has received much attention since MacColl pointed out that it takes no approximate time \cite{1LA,2EH,3HG}. The question about how long it takes an electron to tunnel through a potential barrier is still replete with controversy. The debates center around the definition of tunneling time and its exact physical meaning. Hartman calculated the group delay or phase time and found that the the group delay $\tau_{g}$ becomes constant while the barrier length increases \cite{4TE}. Thus, with a wider barrier, superluminal group velocities can be observed. Recently, Winful proposed that the group delay in tunneling represents a lifetime of stored energy escaping through both sides of the barrier, not a transit time \cite{3HG,5HG,6HG}. However, unlimited group velocity is not a meaningful concept in tunneling and does not imply superluminality.

In the meantime, many optical and acoustic experiments have been carried out to determine tunneling time. Steinberg et al. \cite{7AM} measured the time delay for a photon to tunnel across a  one-dimensional photonic band-gap material. They reported that the group velocity for single-photon tunneling is about $1.7c_{0}$, where $c_{0}$ is  the speed of light in vacuum. The authors attributed superluminality to the fact that the wavepacket may be reshaped in the tunneling process. Longhi et al. \cite{8SL} used a relatively long optical pulse (380ps) and conformed that there is no distortion in the the tunneling process. Even though the exact physical meaning of these experimental results is controversial, there is no doubt that there is a finite duration for the photon or phonon tunneling process. However, few direct experiments have shown that there is finite tunneling time in the quantum particle tunneling process.

There are some difficulties in the measurement of the tunneling time of electrons. For instance, the coherence time of electrons must be long enough to ensure that the tunneling process is an elastic transport. It seems that the graphene is an ideal candidate material for this. The reported electron mobility in graphene at room temperature is in excess of 15,000 $cm^{2}V^{-1}s^{-1}$, and the limit of electron mobility at room temperature  is about 200,000 $cm^{2}V^{-1}s^{-1}$ in theory \cite{9KS,10AK,11AH}. To measure the tunneling time, a high time-resolved technique must be used, so the injected electrons must be generated using ultrafast laser beams (e.g., generated via coherent one- and two-photon absorption  \cite{12RA,13EJ,13HZ}), or the barrier must be controlled by the ultrafast laser beams \cite{14JT,15JT}. Furthermore, because of the existence of zero-point fields, the influences of electromagnetic fields on the tunneling time must be carefully treated. When the electromagnetic fields are included, the barrier is time dependent.  The situation is quite different when a time-dependent potential barrier is taken into account. For instance, for an opaque rectangular barrier with a small time-dependent modulation, the traversal time of B\"{u}ttiker and Landauer is proportional to the barrier width $D$ \cite{16MB}.

\begin{figure}[t]
\includegraphics[width=0.98\columnwidth,clip]{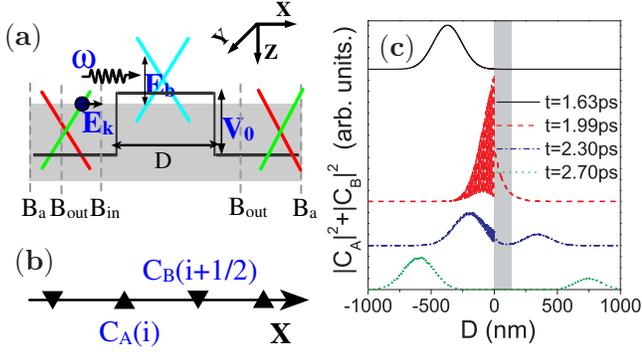}
\caption{(color online). (a) Schematic of the scattering of Dirac
electrons by a square potential irradiated by an electromagnetic field. $B_{a}$, $B_{in}$, and $B_{out}$ denote the absorbing boundary, incident  boundary, and output boundary, respectively. (b) Schematic of the one-dimensional Yee lattice in graphene. (c) The evolution of a wave packet through a barrier with pump intensity $I_{\omega}=20$ MW/cm$^{2}$, $\Delta_{0}=5meV$, $D=150$ nm, and $V_{0}=400$ meV. The light grey shows the barrier area.}%
\label{fig1}%
\end{figure}

In this Letter, we investigate the influences of electromagnetic fields on the tunneling time of Dirac electrons in graphene. We consider a rectangular potential barrier with height $V_{0}$ and width $D$ [see Fig. 1 (a)]. The Fermi level lies in the valence band in the barrier and in the conduction band outside the barrier.  A $Y$-direction polarized laser beam is propagated perpendicular to the layer surface with a detuning $\Delta_{0}=E_{b}-\hbar \omega$. We choose $\Delta_{0}>0$ to ensure that there is no interband absorption inside the barrier, and $\hbar \omega \ll 2 E_{k}$ to guarantee that the influence of the optical field outside the barrier can be neglected. In a single layer graphene, a perfect transmission through a potential barrier in the normal direction is expected, and there is no Hartman effect \cite{17MI,17MI_B,17MI_C,18ZH}. However, if the barrier is irradiated by a intense optical field, the conduction band and valence band is mixed, the chiral symmetry of the Dirac electrons is broken, and the perfect tunneling is strongly suppressed \cite{15JT}. Thus,  the Hartman effect may appear if the electromagnetic field is included.

In order to study such a time-dependent scattering process, we employ the finite-difference time-domain (FDTD) method to solve the time-dependent Dirac equation numerically. In the FDTD method, the time-dependent Dirac equation in $K$ point is replaced by a finite set of finite differential equations

\begin{subequations}
\begin{align}
C_{A}^{k+1/2}(i)&\left[\frac{1}{\Delta t}-\frac{V_{0}(i)}{2i}\right] =\left[\frac{%
1}{\Delta t}+\frac{V_{0}(i)}{2i}\right]C_{A}^{k-1/2}(i) \nonumber \\%
&-\left[ \frac{ v_{F}}{\Delta x}-\frac{V_{12}^{k}(i+1/2)}{2i}\right]
C_{B}^{k}(i+1/2)\nonumber\\%
&+\left[ \frac{v_{F}}{\Delta x}+\frac{V_{12}^{k}(i-1/2)}{2i}\right]
C_{B}^{k}(i-1/2),\label{fdtda}
\end{align}%
\begin{align}
C_{B}^{k+1}&(i+1/2)\left[ \frac{1}{\Delta t}-\frac{V_{0}(i+1/2)}{2i}\right] =%
\left[ \frac{1}{\Delta t}+\frac{V_{0}(i+1/2)}{2i}\right]\times\nonumber \\& C_{B}^{k}(i+1/2)%
-\left[ \frac{v_{F}}{\Delta x}-\frac{V_{21}^{k+1/2}(i+1)}{2i}\right]
C_{A}^{k+1/2}(i+1)\nonumber \\
&+\left[ \frac{v_{F}}{\Delta x}+\frac{V_{21}^{k+1/2}(i)}{2i}%
\right] C_{A}^{k+1/2}(i),\label{fdtdb}
\end{align}%
\end{subequations}
where $(i,k)=(i\Delta x, k\Delta t)$  denotes the grid of point of
the space [see Fig .1(b)], $\mathbf{\Psi }\left( \mathbf{r}, t\right)=[C_{A}(\mathbf{r}, t),C_{B}(\mathbf{r},t)] $ is the wave function of Dirac electrons, $v_{F}\approx 10^{6}m/s$ is the Fermi velocity, $V_{0}(\mathbf{r})$
is the height of the potential barrier, and $V_{12}$ and $V_{21}$ are the matrix elements of the interaction Hamiltonian $\mathbf{H}_{int}$ \cite{13EJ}
\begin{equation}
\mathbf{H}_{int}=-\hbar ev_{F}\left( A_{x}\sigma _{x}+A_{y}\sigma
_{y}\right) =\hbar \left(
\begin{array}{cc}
0 & V_{12} \\
V_{21} & 0%
\end{array}%
\right),
\end{equation}
where $e$ is the electron charge and $(A_{x},A_{y})$ are the vector
potentials of the electromagnetic field.

Thus, by numerically solving Eq. (\ref{fdtda}) and Eq. (\ref{fdtdb}) directly in the time domain, we can demonstrate the propagation of a wave packet through a barrier in real time. For computational stability, we choose the space increment  $\Delta x=0.1$ nm and the time increment  $\Delta t=5\times10^{-5}$ ps. As an example, the evolution of a wave packet through a barrier is shown in Fig. 1(c). At the input boundary $B_{in}$, a Gaussian electronic wave packet $C_{A}=C_{B}=\exp \left[ -\frac{4\pi (t-t_{0})^{2}}{\tau ^{2}}\right]$ is injected. For the convenience of demonstration, a short pulse is used: the peak position $t_{0}=1.2$ ps, and the pulse width $\tau=0.8$ ps. From Fig. 1(c), we find that when the sample is irradiated by an intense nonresonant laser beam, a reflected wave packet appears, and the perfect transmission is suppressed.

Under an intense optical field, the light-induced band shift can also create a dynamic gap at small detuning. As is seen in traditional quantum tunneling (e.g., the Hartman effect), the tunneling time of Dirac electrons should be constant while the barrier width increases. To verify this, we studied the group delay (i.e., the delays of the peak of the reflected and transmitted pulse) of Dirac electrons through a potential barrier generated by an intense light beam. In order to reduce the distortion, a relatively long plus is used: the peak position $t_{0}=5.0$ ps, and the pulse width $\tau=3.3$ ps.

\begin{figure}[t]
\includegraphics[width=0.98\columnwidth,clip]{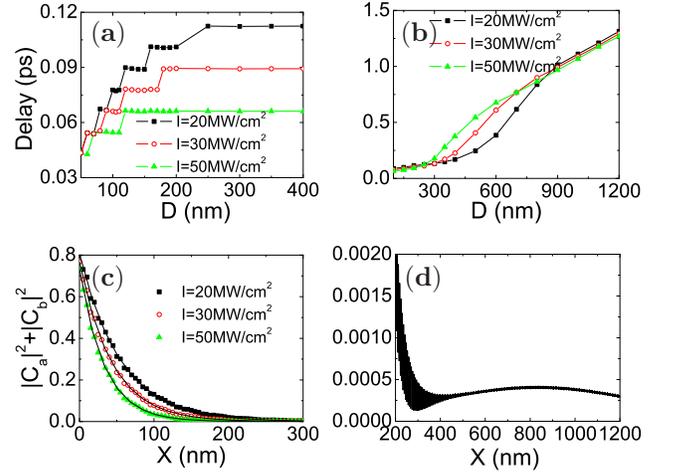}
\caption{(color online). (a) Group delay for the reflected wave
packet and (b) group delay for the transmitted wave packet as a function
of barrier width $D$ at different pump intensities for
$\Delta_{0}=5$ meV. (c) Snapshots of the spatial probability density
distributions on the left side of the barrier for different pump
intensities. The solid lines are the best fit using exponential
functions. (d) Snapshots of the spatial probability density
distribution on the right side of the barrier for $I_{\omega}=50$
MW/cm$^{2}$.}
\label{fig3}%
\end{figure}

Fig. 2(a) shows the group delay for the reflected wave packet as a
function of barrier width $D$ for different pump intensities.  As seen in traditional quantum tunneling, the group delay is saturated by increasing the barrier length. For $I_{\omega}=20$ MW/cm$^{2}$,
$I_{\omega}=30$ MW/cm$^{2}$,  and $I_{\omega}=50$ MW/cm$^{2}$, the
corresponding  saturated delays are about $0.112$, $0.0893$, and
$0.0662$ ps, respectively. The delay can also be explained by the
tunneling depth or dwell time. As shown in Fig. 2(c), the spatial
probability density distributions in the barrier are well fitted by
using the exponential function $F=A_{0}e^{-x/x0}$. For
$I_{\omega}=20$ MW/cm$^{2}$, $I_{\omega}=30$ MW/cm$^{2}$, and
$I_{\omega}=50$ MW/cm$^{2}$, the corresponding  tunneling depths
$x_{0}$ are $55$, $44$, and $33$ nm, respectively. In this way, we
extract that the corresponding tunneling delays $t_{0}=2x_{0}/v_{F}$ are
about $0.11$, $0.088$, and $0.066$ ps, respectively. These results
are consistent  with the saturated delay obtained in Fig. 2(a).
Another interesting phenomenon is that the group delay for the
reflected wave packet takes on the quantized values $\tau=n T_{1}$
$(n=1,2,3...)$. We find that $T_{1}\approx0.011$ ps is equal to the
period of the optical field. The quantized  group delay is
therefore caused by the optical modulation of the reflected wave
packet.

The case is quite different for the transmitted wave packet. As
shown in Fig. 2(b), when the barrier width $D>900$ nm, the group
delay increases linearly with increasing barrier width. The group
velocity is about  $10^{6}$ m/s and is the same as the Fermi
velocity of Dirac electrons in graphene. This result can be
explained by the time-dependent  potential barrier. In a
time-dependent  potential barrier, the energy storage depends on
time. Variations in energy storage will lead to an extra leakage
current. The dynamic gap caused by the optical field is therefore not a complete
gap, and the probability density distribution on the right side of
the barrier is no longer exponential decay [see Fig. 2(d)]. Since
the amplitude of the extra leakage current is quite small, it has
little effect on the reflected wave packet but still determines the
group delay for the transmitted wave packet at large barrier widths.
For small barrier widths, since the tunneling current is much larger
than the extra leakage current, the time delays of the reflected and
transmitted wave packets are equal. From Fig. 2(b), we see that the
amplitude of the extra leakage current strongly depends on the pump
intensity. Specificially, for $I_{\omega}=50$ MW/cm$^{2}$, the group
delay increases linearly with increasing barrier width when $D>700$
nm, but for $I_{\omega}=20$ MW/cm$^{2}$ the linearly increasing
delay appears when $D>900$ nm. However, the group velocity of the
extra leakage current is independent of pump intensity. Thus,
even with a quite small time-dependent modulation, no "Hartman
effect" occurs.

\begin{figure}[t]
\includegraphics[width=0.98\columnwidth,clip]{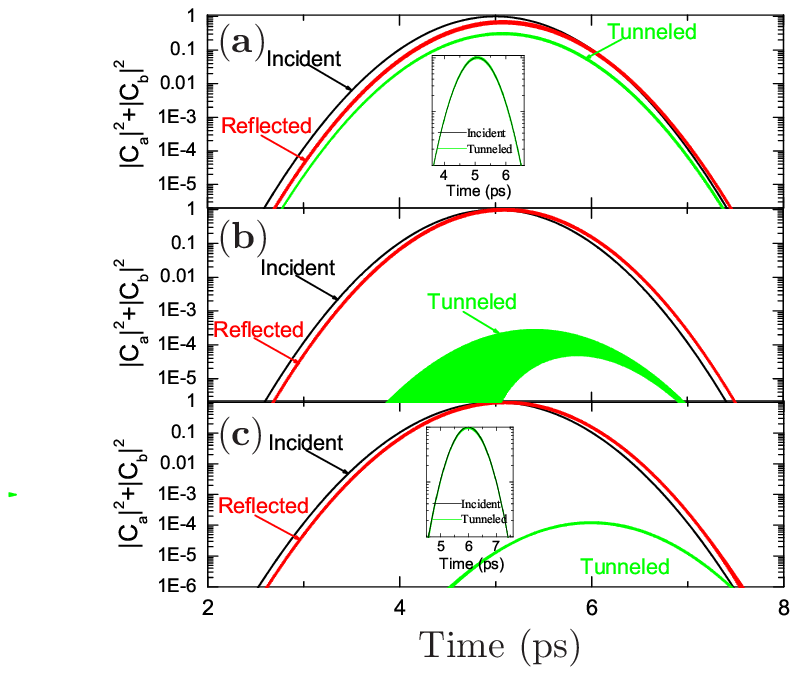}
\caption{(color online). Incident (black lines), tunneled (green lines), and
reflected (red lines) pulses with pump intensity $I_{\omega}=30$ MW/cm$^{2}$  for the different barrier width (a) $D=100$ nm, (b) $D=500$ nm, and (c) $D=900$ nm. The insets shows the normalized tunneled pulse overlaid with the incident pulse.}
\label{fig3}%
\end{figure}

The existence of current leakage in a time-dependent barrier can also be confirmed by the shape of tunneled pulses. In a thin barrier, the amplitude of the tunneling current is much larger than the amplitude of the extra leakage current induced by the time-dependent modulation. The tunneling time of the transmitted and reflected wave packet is equal [see Fig. 3(a)], and the distortion is quite small [see the inset of Fig. 3(a)]. When the thickness of the barrier increases, the tunneling rates decrease rapidly. In a thick barrier, the amplitude of the extra leakage current is comparable with that of the tunneling current. Since the tunneling times of the extra leakage current and the tunneling current are different, a serious distortion can be found in the tunneled pulse [see Fig. 3(b)]. However, if the width of the barrier is large enough (e.g., $D=900$ nm),  the tunneling  rate is very small, and the extra leakage current is the main contributor to the tunneled pulse. Thus, the tunneling times of the transmitted and reflected wave packets are different [see Fig. 3(c)], and there is no distortion on this scale [see the inset of Fig. 3(c)].

The tunneling delay for a time-dependent potential barrier might
help us to understand quantum tunneling. For example,
traditional definitions of tunneling time are unsuitable for the
system we studied. In the traditional definitions, the tunneling
time of the transmitted and reflected wave packets are equal for a
symmetrical barrier. More importantly, if the quantum fluctuation or
the zero-point field is included, all potential barriers are time
dependent.

In conclusion, we have calculated  the influence of optical fields of
chiral tunneling time in graphene using the FDTD method. We find that the group delay of the reflected packet is also saturated as the barrier width increases. However, the delay increases linearly with barrier length for the transmitted wave packet. This peculiar tunneling effect is attributed to current leakage in a
time-dependent barrier generated via the optical field. Thus, the zero-point field may have an important influence on the tunneling time of electrons, and should be carefully treated.  These unique transport properties of dressed Dirac electrons in graphene might be important in the understanding of quantum tunneling.

This work was supported by the NSFC Grant Nos. 10904059 and
10904097, the NSF from Jiangxi Province Nos. 2008GZW0003 and
2009GQW0017, the Open Research Fund of State Key Laboratory of
Millimeter Waves No. K200901, and the Program for Innovative
Research Team in Jiangxi Province.

\end{document}